\documentclass{ws-mpla}
\usepackage[super]{cite}
\usepackage{graphicx}
\begin{document}
\markboth{Snehasish Bhattacharjee}
{Magnetic effects on the inner discs around neutron stars}

\catchline{}{}{}{}{}

\title{Effects of magnetic field on the radiation pressure dominated discs around neutron stars}

\author{\footnotesize Snehasish Bhattacharjee\footnote{
snehasish.bhattacharjee.666@gmail.com}}

\address{Department of Astronomy, Osmania University, Hyderabad-500007,
India}

\maketitle

\pub{Received (3 January 2020)}{Revised (25 January 2020)}

\begin{abstract}
We supplement the analytic solution obtained by Matthews O. M., et. al., 2005, MNRAS, \textbf{356}, 66 to investigate the steady-state structure of radiation pressure dominated disc under the influence of a stellar magnetic field which deploys a torque. The solutions converge to the non-magnetic Shakura N. I., Sunyaev R. A., 1973, A\& A, \textbf{24}, 337 form when magnetic field of the star tends to zero and also at large radii. Effects of varying the mass accretion rates and the spin period of a typical neutron star on the disc parameters are presented. We further report that the presence of a magnetic correction term $k$ reduces the radial extent upto which radiation pressure and electron scattering continue to be the major source of pressure and opacity respectively. We also report that magnetic effects amplify the viscous timescale several times in the inner disc.

\keywords{accretion disc; neutron stars; magnetic field}
\end{abstract}

\ccode{PACS Nos.: 97.60.Jd;97.10.Gz}

\section{Introduction}
Shakura \& Sunyaev \cite{ss} presented a detailed solution for the structure of thin accretion discs around black holes. The solutions contain expressions for disc scale-height, surface density, pressure, temperature, viscosity and opacity. The solutions depend on mass accretion rate, mass of the central star, radius of the star and radial extent. Additionally the solutions also depend on the dimensionless parametrization of viscosity called $\alpha$ parameter. \\
But Shakura-Sunyaev solution does not include the effects of torque from the central star. Since accretion discs are also found near magnetic stars, Matthews, et al., \cite{matt} reconstructed the Shakura-Sunyaev disc solutions in the case of magnetic stars. A stellar magnetic field contributes to the transfer of angular momentum from the inner parts of the disc to the outer parts and can also govern the constitution of the disc and spin evolution of the star \cite{AM1996,BC1998}. These effects are significant in cataclysmic variables \cite{S2002}, X-ray binaries \cite{PR1972,Rom2003} and young stars \cite{K1990,ACT1999}. The solution obtained in \cite{matt} is continuous throughout the disc and is applicable to a wide range of opacity models but the solution is not accurate in the very inner parts of the disc where the disc suffers from radiation pressure.\\
Radiation pressure starts to dominate in the very inner parts of the disc when the mass accretion rates is high and as a consequence the opacity is mainly due to electron scattering and in such cases the disc is thermally and viscously unstable \cite{LE1974,SS1976,SH1975}.\\ 
Thus we extend the work of Matthews \textit{et. al.,} \cite{matt} to study the innermost regions of a steady thin disc and investigate the structure of the disc dominated by radiation pressure. We also provide analytic expressions separating distinct disc regions in magnetic case. We report that except for very high mass accretion rates $(\dot{M_{16}}>1)$ the inner disc is absent and hence Kramers opacity can be valid throughout the disc. The effects of varying spin periods of the central star on the disc parameters are analyzed with respect to the non-magnetic Shakura-Sunyaev solution. Our solution converges to Shakura-Sunyaev solution when the magnetic field of the star tends to zero and also at large radii similar to the solution obtained in \cite{matt}. We also report necessary modification to viscous timescales ($t_{vis}$) in both inner and outer discs due to the presence of a magnetic correction term. Such modifications increases $t_{vis}$ by many folds in the inner disc while for the outer disc $t_{vis}$ is not significantly affected.\\
The paper is organized as follows: In Section II we provide the basic assumptions and theory needed to formulate the disc solutions and report analytic solutions of the disc parameters and study their behavior graphically. In Section III we provide analytic expressions of boundary radii separating different disc regions. In Section IV we furnish the modification of viscous timescale in both inner and outer discs. Finally in Section V we present our conclusions.  

\section{Disc Solutions}
\subsection{Essential Ingredients}
The derivations of disc parameters involve nine equations which we derive following the prescriptions given in \cite{matt} and \cite{frank}.\\
First we require an equation which encapsulates the structure of the disc in the vertical direction. Since there is no flow of matter in this direction, hydrostatic equilibrium is maintained and thus the scale-height $H$ of the disc for a thin disc approximation can be written as
 \begin{equation}\label{1}
H=\frac{ c_{s}}{\omega}
\end{equation}
where $c_{s}$ is the local sound speed, $\omega = \sqrt{\frac{G M_{\ast}}{R^{3}}}$ is the angular velocity of the disc material at radial distance $R$ from the accretor in the plane of the disc, $G$ is the universal gravitational constant and $M_{\ast}$ the mass of the central object. Furthermore, \eqref{1} also implies \cite{frank} 
\begin{equation}
c_{s}\ll \left( \frac{G M_{\ast}}{R}\right)^{1/2} 
\end{equation}
i.e. the local Keplerian velocity for a thin disc should be highly supersonic. As mentioned in \cite{frank}, this illustrates a constrain on the temperature of the disc and finally on the overall cooling mechanism.  \\
For a thin disc, computation of disc structure become simplified owing to the fact that both the temperature and pressure gradients are vertical and thence radial and vertical structures are mainly decoupled \cite{frank}. \\
We now require an equation which gives the speed of sound propagating in the thin isothermal disc. The expression reads
\begin{equation}\label{3}
c^{2}_{s}=\frac{P}{\rho}
\end{equation}
where $P$ represent total pressure acting on the fluid and $\rho$ represents the density of the fluid in the disc. Expression of $\rho$ reads
\begin{equation}\label{4}
\rho = \frac{\Sigma}{H} 
\end{equation}
where $\Sigma$ represents the surface density of the fluid in the disc.\\
Total pressure $P$ is made up of three parts. These are: gas pressure ($P_{gas}$), radiation pressure ($P_{radiation}$) and magnetic pressure ($P_{magnetic}$) and can be expressed as
\begin{equation}
P= P_{gas} + P_{radiation} + P_{magnetic} = \frac{\rho \kappa T_{c}}{\epsilon m_{p}}+ \frac{4 \sigma T^{4}_{c}}{3 c} + \frac{4 \pi}{\mu_{0}} \frac{\mu^{2}}{8 \pi r^{6}}
\end{equation}
where $\sigma$ is the Stefan-Boltzmann constant, $T_{c}$ the temperature in the midplane of the disc ($z=0$), $\kappa$ the Boltzmann constant, $m_{p}$ and $\epsilon$ denote respectively the atomic mass unit and mean molecular weight in the disc, $\mu_{0}$ being the vacuum permeability and $\mu$ the magnetic moment of the star which is given by $\mu= B_{\ast}\times R_{\ast}^{3}$. Here, $B_{\ast}$ is the magnetic field of the star and $R_{\ast}$ represents the stellar radius. In the inner disc, radiation pressure dominates over gas pressure and hence we neglect the contributions of gas pressure in our derivations. Hence the total pressure ($P$) in the inner disc for the magnetic case becomes 
\begin{equation}
P =  \frac{4 \sigma T^{4}_{c}}{3 c} + \frac{4 \pi}{\mu_{0}} \frac{\mu^{2}}{8 \pi r^{6}}
\end{equation}
For a highly magnetized star with magnetic fields of the order of $B\gtrsim 10^{12} G$, the disc gets disrupted in the very inner regions. However, modeling such interactions is exceedingly complex and therefore in \cite{frank} a simple case was considered where the magnetic field interacts and disrupts the flow which is quasi-spherical in nature. Indeed it was shown in \cite{frank} that for such high magnetic fields, the spherically symmetric matter flow is disrupted by the magnetic pressure ($P_{magnetic}$) at the radius $r_{RM}$, where $P_{magnetic}$ exceeds the ram pressure $P_{ram}$. By setting $P_{magnetic}$  equal to ram pressure ($P_{ram} = \rho v^{2}$, where $v$ is the velocity of fluid), the expression of $r_{RM}$ reads \cite{frank}
\begin{equation}
r_{RM} = 5.1\times 10^{8} M_{1}^{-1/7} \dot{M}_{16}^{-2/7} \mu _{30}^{4/7}
\end{equation}
where $M_{1}=\frac{M_{\ast}}{M_\odot}$ is the mass of the star in units of solar mass and $\dot{M}_{16}=\frac{\dot{M}}{10^{16}(gm/s)}$ is the mass accretion rate in units of $10^{16}(gm/s)$ and $\mu_{30}$ is represented in units of $10^{30}$ G $cm^{3}$. For a neutron star with $M_{1}=1.4$, $R_{\ast}\simeq 10$ km and $B\simeq10^{12}$G, the radius $r_{RM} \sim 5 \times 10^{3}$km. At such large distances from the accretor the contributions of $P_{radiation}$ to the total pressure is negligible. Hence, our analysis is valid and useful only when the mass accretion rate is high, i.e. $\dot{M}_{16}>50$ and $B\sim 10^{8}-10^{10}G$. Under these circumstances, $r_{RM}\sim R_{\ast} $ and $P_{radiation}$ becomes the major source of pressure in the very inner regions of the disc. Hence, the total pressure under these set of conditions to a good approximation can be written as   
\begin{equation}\label{2}
P=\frac{4 \sigma T^{4}_{c}}{3 c}
\end{equation}

We now need an equation for the conservation of energy in the disc. The gravitational potential energy of the infalling material gets converted into heat energy and radiation. Additionally, for very high magnetic fields, the amount of magnetic energy generated is substantial and sometimes comparable to the gravitational power \cite{kool}. Nonetheless, it has been argued that this energy mainly dissipates in the corona because dissipation inside the disc is negligible owing to the low Alfv\'en speed \cite{kool,galeev}. Hence  we obtain \cite{frank}
\begin{equation}\label{5}
\frac{4T^{4}_{c}\sigma}{3\tau}=\frac{9}{8}\nu\Sigma\frac{GM_{\ast}}{R^{3}}
\end{equation} 
where $\tau$ is the opacity of the fluid in the disc. In the inner disc the gas is fully ionized with a temperature $T \gtrsim 10^{4}$K and the opacity is mainly due to electron scattering which is given by \cite{ss}  
\begin{equation}\label{6}
\tau=\Sigma\frac{\sigma_{t}}{m_{p}}; \hspace{0.25in} \text{ where} \hspace{0.1in} \frac{\sigma_{t}}{m_{p}} =0.4 cm^{2}gm^{-1}
\end{equation}
Next, the viscosity $\nu$ of the fluid in the disc can be expressed as \cite{ss}
\begin{equation}\label{7}
\nu = \alpha c_{s} H
\end{equation}
Where $\alpha< 1$ is a constant. In addition to the above equations, an expression for $\nu$ $\Sigma$ in the magnetic case is also required which is obtained from Matthews \textit{et. al.,} (2004) \cite{matt} and reads \begin{equation}\label{8}
\nu\Sigma=\frac{\dot{M}}{3\pi}d^{4}k
\end{equation}
where the parameter $k$ reads
\begin{equation}\label{9}
k=1-\frac{\beta}{\dot{M}}\frac{\pi h}{R^{\gamma}(\gamma - 2) d^{4}}
\end{equation} 
where $\gamma = \frac{7}{2}$ for a fully magnetized disc \cite{matt} and $\beta$ is known as magnetic parameter and reads \cite{matt}
\begin{equation}\label{10}
\beta= \frac{\mu^{2}}{2\pi\sqrt{G M_{\ast}}}
\end{equation}
The parameter $d$ is expressed as
 \begin{equation}\label{11}
d=\left[ 1-\left( \frac{R_{t}}{R}\right) ^{\frac{1}{2}}\right] ^{\frac{1}{4}}
\end{equation}
where $R_{t}$ represents truncation radius of the disc in the presence of a stellar magnetic field. Finally $h$ is expressed as
\begin{equation}\label{12}
h=\left( \frac{R}{R_{co}}\right) ^{\frac{3}{2}}\left [1-\left( \frac{R_{t}}{R}\right) ^{(2-\gamma)}\right]
-\left( \frac{\gamma -2}{\gamma - \frac{1}{2}}\right) \left [1-\left( \frac{R_{t}}{R}\right) ^{(\frac{1}{2}- \gamma)}\right]
\end{equation}
The corotation radius $R_{co}$ is the radius where the plasma pressure equals the static magnetic field pressure. At this radial distance from the accretor, the plasma gets frozen in the magnetic field and rotates rigidly with the angular velocity of the accretor. This radius is a function of the spin period of the star $P_{spin}$ and the mass of the accretor $M$ and can be expressed as \cite{frank}
\begin{equation}\label{13}
R_{co}[cm]= 1.5\times10^8\left[\frac{P_{spin}}{1 sec}\right]^{\frac{2}{3}}\left[\frac{M_{\ast}}{M_\odot}\right]^{\frac{1}{3}} 
\end{equation}
In presence of a stellar magnetic field the disc gets truncated at a radius which is a function of $\dot{M}$, $P_{spin}$, $R_{\ast}$ and $\beta$ \cite{matt} and the expression of $R_{t}$ reads 
 \begin{equation}\label{14}
\frac{R_{t}}{R_{co}}=\left[ \frac{Q}{2}\left( \sqrt{\frac{4}{Q}+1}-1\right) \right] ^{\frac{2}{3}}
\end{equation}
where $Q$ is defined as
 \begin{equation}\label{15}
Q=2\pi R_{co}^{-\gamma}\left( \frac{\beta}{\dot{M}}\right) 
\end{equation}

\subsection{Expressions of Disc Parameters}
Following a procedure similar to that in \cite{matt} we obtain the expressions of disc parameters for the inner disc dominated by radiation pressure in presence of a stellar magnetic field which exerts a torque on the disc.\\
Substituting equation ($2$) into equation ($5$), we obtain\begin{equation}\label{16}
\left( \frac{4\sigma}{3\tau}\right) \left( \frac{3c}{4\sigma}\right) P=\frac{9}{8}\nu\Sigma\frac{GM_{\ast}}{R^{3}}
\end{equation}
Substituting equations (\ref{1}, \ref{3}, \ref{4}, \ref{6} \& \ref{7}) into equation (\ref{16}) we get,
\begin{equation}\label{17}
H \omega^{2} c \frac{m_{p}}{\sigma_{T}}=\frac{9}{8} \omega^{2} \frac{\dot{M}}{3 \pi} d^{4} k
\end{equation}
Introducing dimensionless quantities \\$M_{1}=\frac{M_{\ast}}{M_\odot}$, \hspace{0.1in} $R_{10}=\frac{R}{10^{10}cm}$, \hspace{0.1in} $\dot{M_{16}}=\frac{\dot{M}}{10^{16}(gm/s)}$\\
we finally obtain the scaleheight ($H$) of the disc as \begin{equation}\label{18}
H\left[ cm\right] = 1.6\times10^4 \dot{M_{16}} d^{4} k 
\end{equation} 
The remaining disc parameters are obtained easily and are collected below
\begin{equation}\label{19}
c_{s}\left[\frac{cm}{s}  \right] = 183 \dot{M}_{16} M_{1}^{1/2} R_{10}^{-3/2} d^{4} k
\end{equation}
\begin{equation}\label{20}
\nu\left[\frac{cm^{2}}{s} \right] =2.9\times10^5 \alpha \dot{M}_{16}^{2} M_{1}^{1/2} R_{10}^{-3/2} d^{8} k^{2}
\end{equation}
\begin{equation}\label{21}
\Sigma\left[\frac{gm}{cm^{2}} \right] =3.6\times10^8 \alpha^{-1} \dot{M}_{16}^{-1} M_{1}^{-1/2} R_{10}^{3/2} d^{-4} k^{-1}
\end{equation}
\begin{equation}\label{22}
V_{r}\left[\frac{cm}{s} \right] =4.4\times10^{-4} \alpha\dot{M}_{16}^{2} R_{10}^{-5/2}M_{1}^{1/2}d^{4}k
\end{equation}
\begin{equation}\label{23}
\tau=1.4\times10^8 \alpha^{-1} M_{1}^{-1/2} R_{10}^{3/2} \dot{M}_{16}^{-1} d^{-4} k^{-1} 
\end{equation}
\begin{equation}\label{24}
\rho \left[ \frac{gm}{cm^3}\right] =2.3\times10^4 \alpha^{-1} \dot{M}_{16}^{-2} M_{1}^{-1/2} R_{10}^{3/2}  d^{-4} k^{-1}
\end{equation}
\begin{equation}\label{25}
T_{c}\left[K \right] = 7.4\times10^5 \alpha^{-1/4} M_{1}^{1/8} R_{10}^{-3/8}
\end{equation}
\begin{equation}\label{26}
P_{rad}\left[ \frac{dyne}{cm^2}\right] =7.6\times10^8 \alpha^{-1} M_{1}^{1/2} R_{10}^{-3/2}
\end{equation}
Our expressions of disc parameters in the magnetic case differ from the non-magnetic Shakura \& Sunyaev solution only in the presence of magnetic correction term $k$ and replacing $R_{t}$ for $R_{\ast}$ everywhere. As reported in \cite{matt} the correction term $k$ vary from unity by an amount that is proportional to the ratio of $\beta$ to the mass transfer rate. Thus if mass transfer rate is high it can overpower the magnetic field of the accretor and the disc will reduce back to non-magnetic Shakura \& Sunyaev form. 
\subsection{\textbf{Graphical Representation}}
\subsubsection{For Varying Spin Period:}
\begin{figure}[h]
\centerline{\includegraphics[width=3.2in]{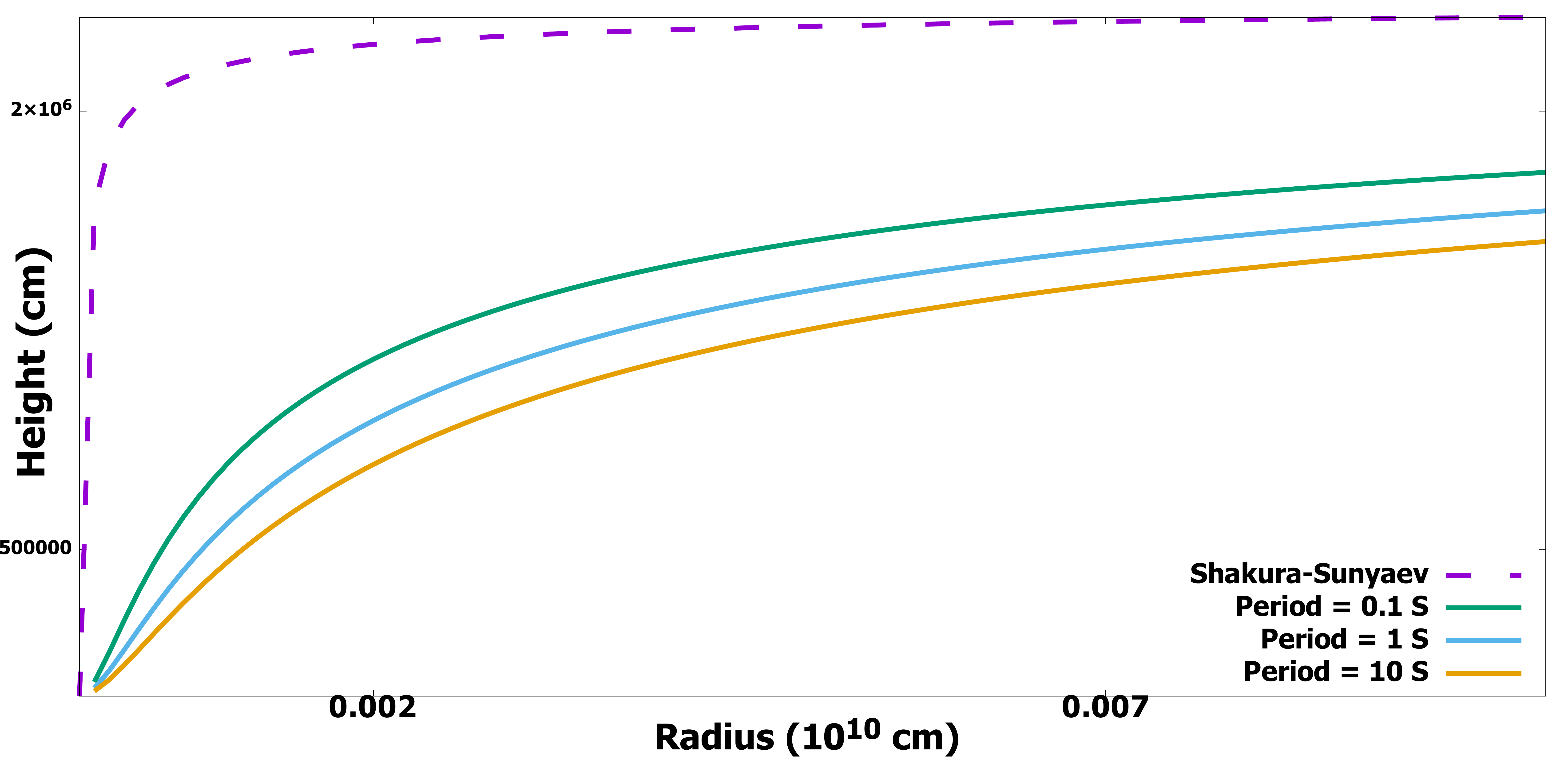}}
\vspace*{8pt}
\caption{Height ($H$) vs radius for different spin periods of a neutron star with $ P_{spin} = 0.1, 1$ and $10$ sec.\protect\label{f1}}
\end{figure}

\begin{figure}[h]
\centerline{\includegraphics[width=3.2in]{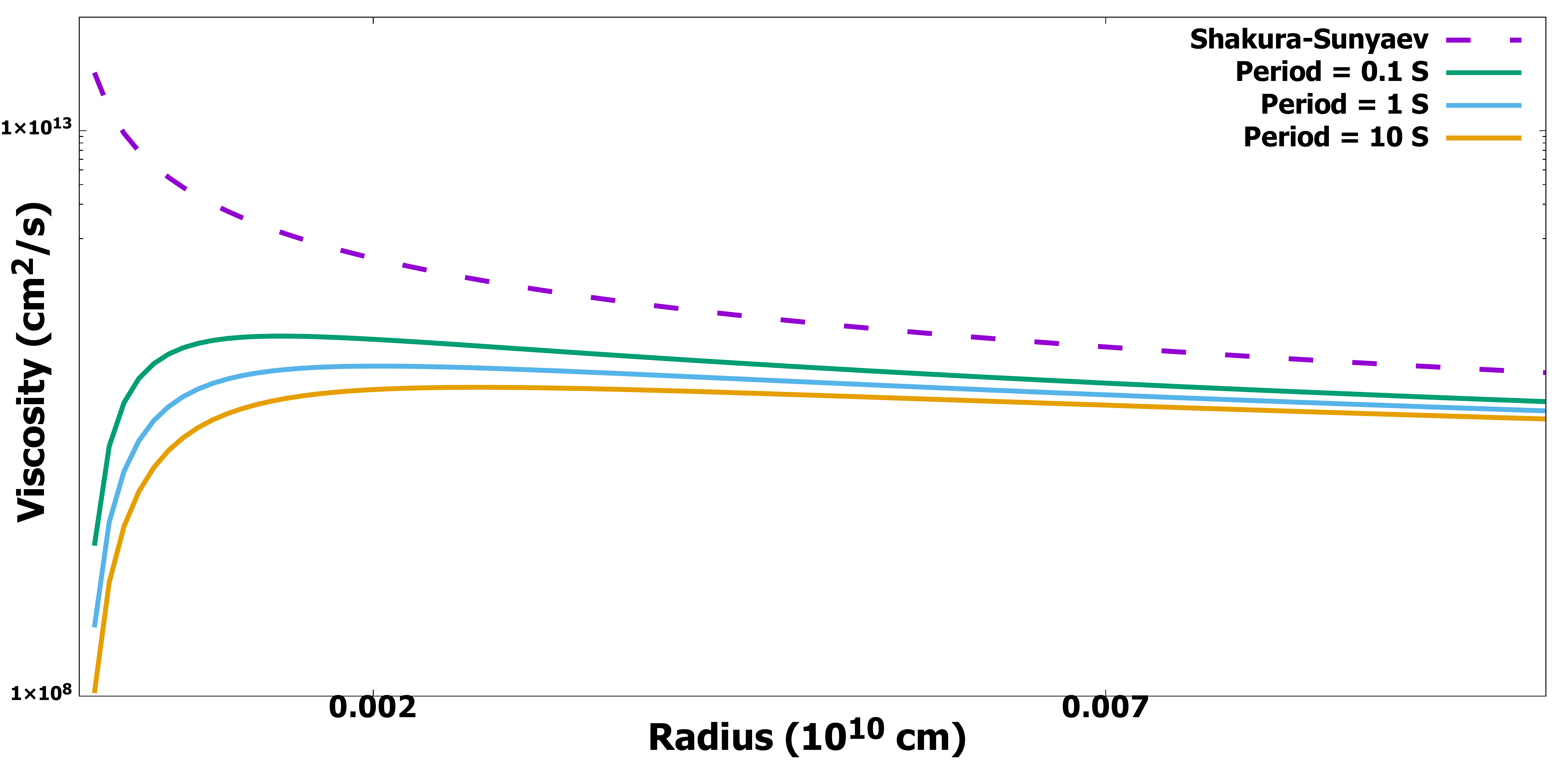}}
\vspace*{8pt}
\caption{Viscosity ($\nu$) vs radius for different spin periods of a neutron star with $ P_{spin} = 0.1, 1$ and $10$ sec.\protect\label{f2}}
\end{figure}

\begin{figure}[h]
\centerline{\includegraphics[width=3.2in]{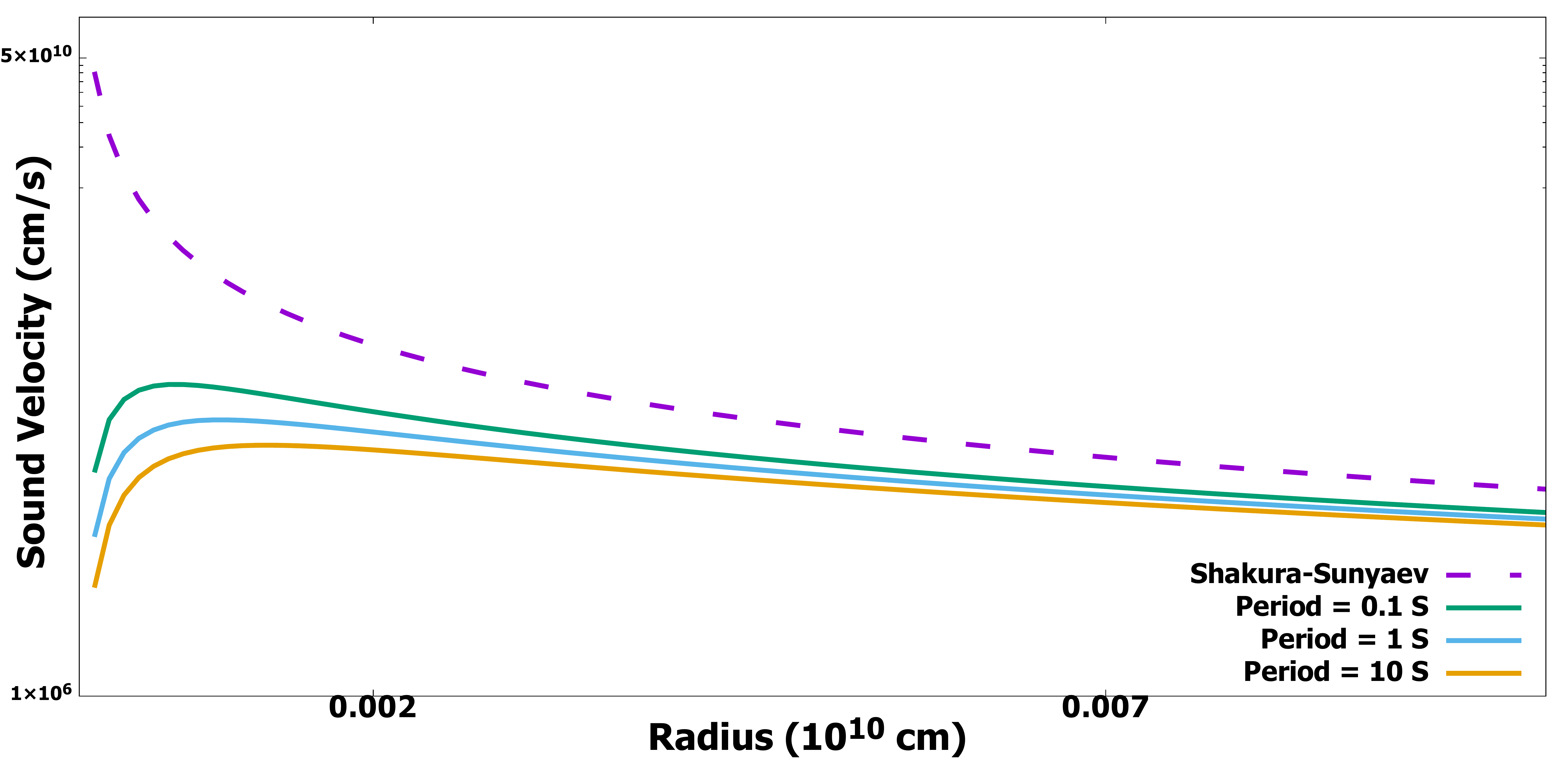}}
\vspace*{8pt}
\caption{Sound speed ($c_{s}$) vs radius for different spin periods of a neutron star with $ P_{spin} = 0.1, 1$ and $10$ sec.\protect\label{f3}}
\end{figure}

\begin{figure}[h]
\centerline{\includegraphics[width=3.2in]{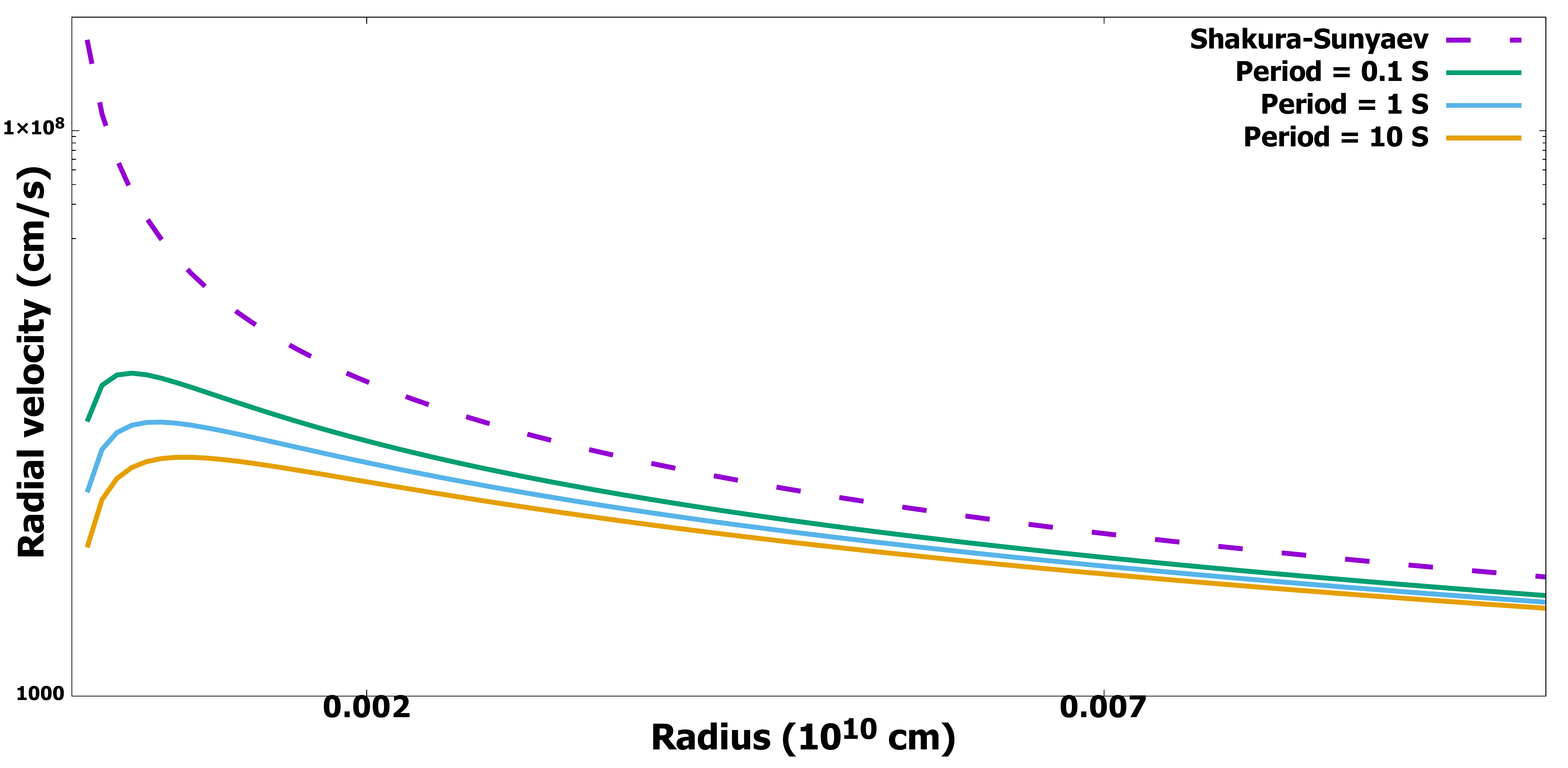}}
\vspace*{8pt}
\caption{Radial velocity ($V_{r}$) vs radius for different spin periods of a neutron star with $ P_{spin} = 0.1, 1$ and $10$ sec.\protect\label{f4}}
\end{figure}

\begin{figure}[h]
\centerline{\includegraphics[width=3.2in]{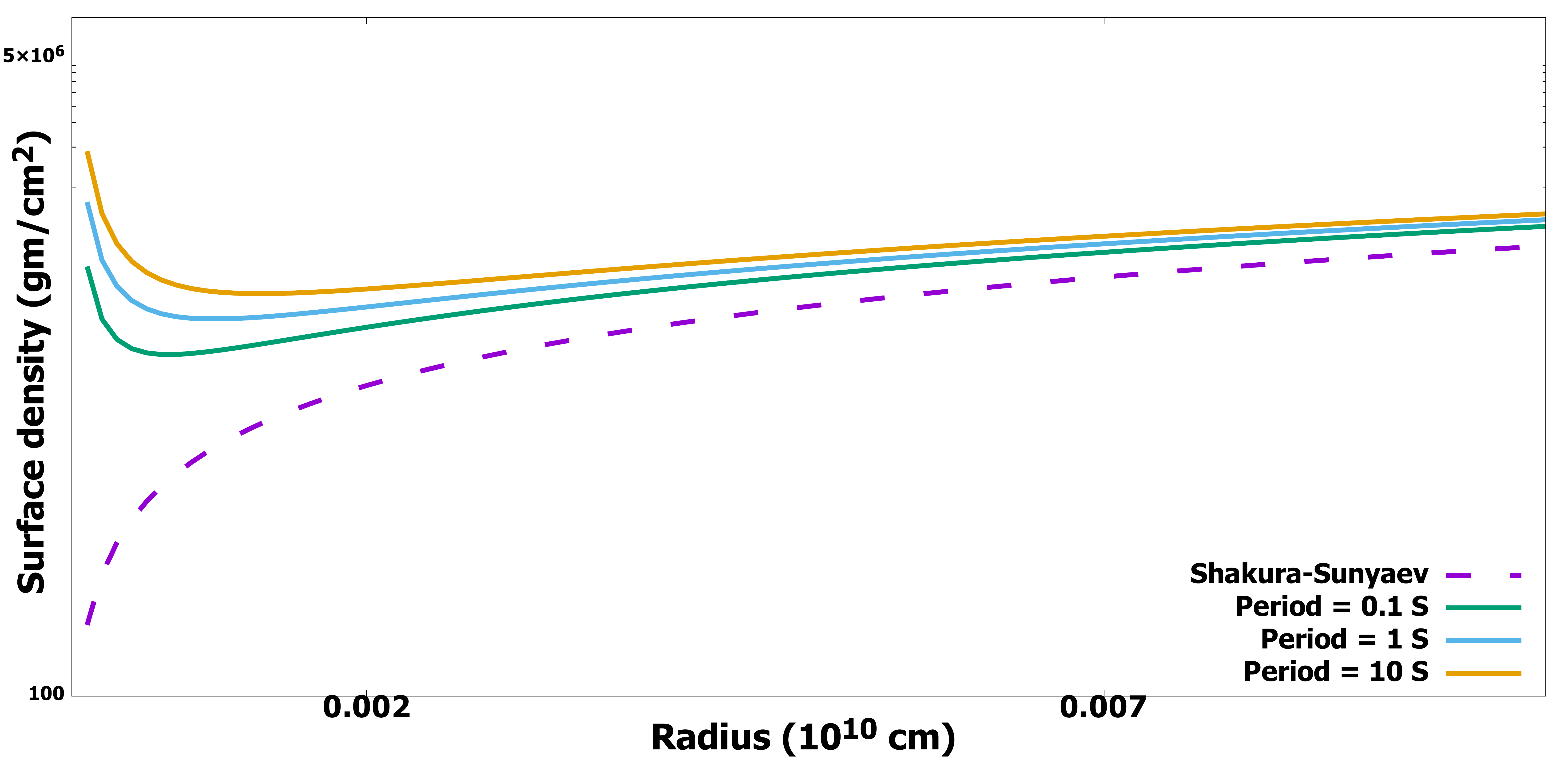}}
\vspace*{8pt}
\caption{Surface density ($\Sigma$) vs radius for different spin periods of a neutron star with $ P_{spin} = 0.1, 1$ and $10$ sec.\protect\label{f5}}
\end{figure}

Eqs. \ref{18} - \ref{22} are graphically shown in Figs \ref{f1}-\ref{f5} for different spin periods of a neutron star ranging from $0.1 - 10$ sec. Stellar mass is set to $M_{\ast} = 1.4 M_\odot$, radius of the star $R_{\ast} = 10^6 cm$, mass accretion rate $\dot{M}= 1.5\times10^{18} gm/s$, alpha parameter takes the value $\alpha = 0.01$ and the magnetic field is set to $10^8 G$. We consider the magnetic field to be low because as we will see in Section \eqref{bd} that for high magnetic fields, the inner disc is cut and hence magnetic term $k$ approaches unity. The solutions are plotted for $R = 0 - 1000$ km. We also show Shakura \& Sunyaev solutions which describes non-magnetic profiles of the disc parameters.


In the inner disc $H$ dependency on $R$ is only due to the presence of $k$ term (equation \ref{18}) which becomes unity when magnetic moment of the star vanishes and hence for non-magnetic stars (usually black holes) the vertical scaleheight is constant throughout the region as seen in Figure \ref{f1}. However in the gas pressure dominated disc, $H$ increases almost linearly with radius \cite{matt}. Since $0\leq k \leq1$ disc becomes thickest for the non-magnetic case ($k=1$). Presence of magnetic field makes the disc thinner. It is to be noted that for Shakura \& Sunyaev profile, truncation occurs at the surface of the star ($R_{\ast}$) while for magnetic stars, truncation occurs further out due to magnetically induced accretion. The disc becomes thicker as spin period of the star decreases, but in any case $H \leq 0.1 R$ and hence thin disc approximation is maintained. \\
Figure \ref{f2} shows profile of viscosity where it is clearly observed that viscosity for non-magnetic case is several orders of magnitude higher than in magnetic cases. This is because $\nu$ is strongly dependent on function $k$ evident from equation (\ref{20}) ($\nu\propto k^{2}$). Therefore viscosity in magnetic cases decreases as spin period increases owing to the fact that $k$ is inversely proportional to the spin period. Surprisingly in the outer gas pressure dominated disc, viscosity is weakly dependent on $k$ term ($\nu \propto k^{3/10}$) and increases steadily with radius ($\nu \propto R^{3/4}$) \cite{matt}.\\ 
Figure \ref{f3} illustrates sound velocity ($c_{s}$) as a function of radius where we see $c_{s}$ for the non-magnetic case is much higher compared to all the magnetic cases in the vicinity of the truncation radius $R_{t}$.  This again is attributed to the presence of magnetic term $k$ in equation \ref{19} which is always less than unity for magnetic stars (usually neutron stars). An interesting thing to note here is that in the radiation pressure dominated discs $H$ and $c_{s}$ are independent of $\alpha$ parameter which can be clearly seen from equations \ref{18} \& \ref{19}.\\
Figure \ref{f4} depicts profiles of radial velocity. Since $V_{r} \propto \dot{M}^{2} R^{-5/2}$, for very high accretion rates ($\dot{M}_{16}\geq 1$) and very near to the truncation radius ($R_{t}$) the Shakura \& Sunyaev profile deviate the most from the magnetic ones. However away from $R_{t}$, the magnetic profiles converge to the Shakura \& Sunyaev solution. It is also seen that $V_{r}$ for the non-magnetic case is clearly higher than all the magnetic cases. $V_{r}$ increases as the spin period ($P_{spin}$) of the central star decreases because an increase in $P_{spin}$ drives $R_{t}$ outwards decreasing $d$ (equation \ref{9}) and ultimately increasing the magnetic term $k$ (equation \ref{10}). From Figure \ref{f3} and \ref{f4} it is also observed that in every case $\frac{V_{r}}{c_{s}}\leq 0.001$.\\ 
Figure \ref{f5} shows surface density ($\Sigma$) as a function of radius. From equation \ref{21} we note that $ \Sigma \propto \dot{M_{16}}^{-1} k^{-1}$ and hence an increase in mass accretion rate diminishes the surface density. However as spin period increases, we see significant enhancement of surface density in view of the the fact that an increase in $P_{spin}$ results in a decrease in $k$ and therefore amplifies the surface density. Interestingly in the gas pressure dominated disc, an increase in mass accretion rate makes the disc denser Matthews \textit{et. al.,} \cite{matt}. 
As reported in \cite{matt} there is a limit to the increase in surface density due to two reasons. Either an increase in $\Sigma$ makes the disc so thick that thin disc approximation becomes obsolete or an ever  increasing surface density results in a thermal-viscous outburst at some time.\\
In the radiation pressure dominated discs, the temperature ($T_{c}$) and pressure ($P_{rad}$) are independent of magnetic correction term $k$ evident from equations (\ref{25}) and (\ref{26}).
\subsubsection{For Varying Mass Accretion Rate:}
\begin{figure}[h]
\centerline{\includegraphics[width=3.2in]{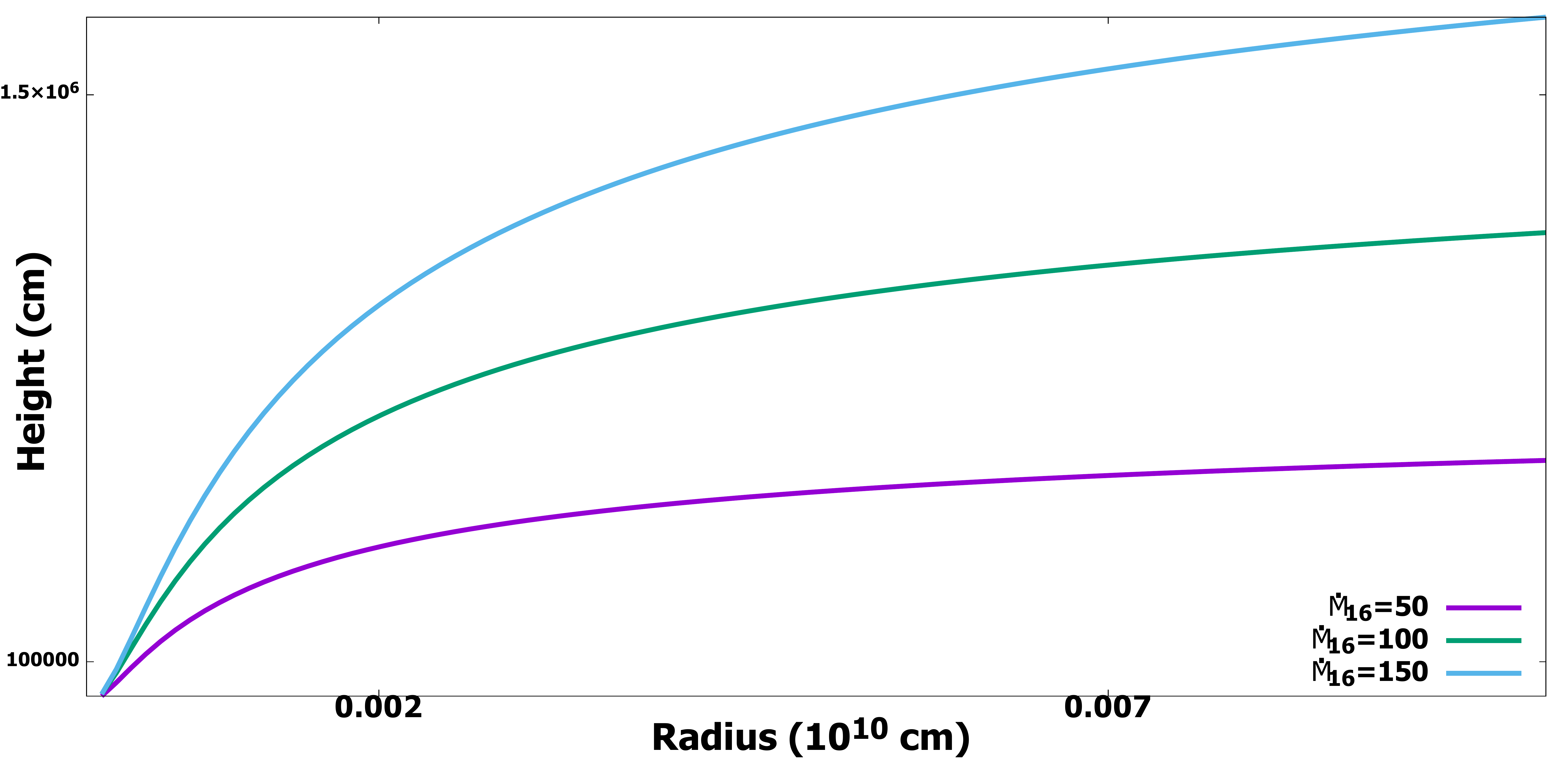}}
\vspace*{8pt}
\caption{Height ($H$) vs radius for varying mass accretion rates ($\dot{M}_{16}$) for the cases $\dot{M}_{16} = 50 , 100$ and $150$.\protect\label{f6}}
\end{figure}

\begin{figure}[h]
\centerline{\includegraphics[width=3.2in]{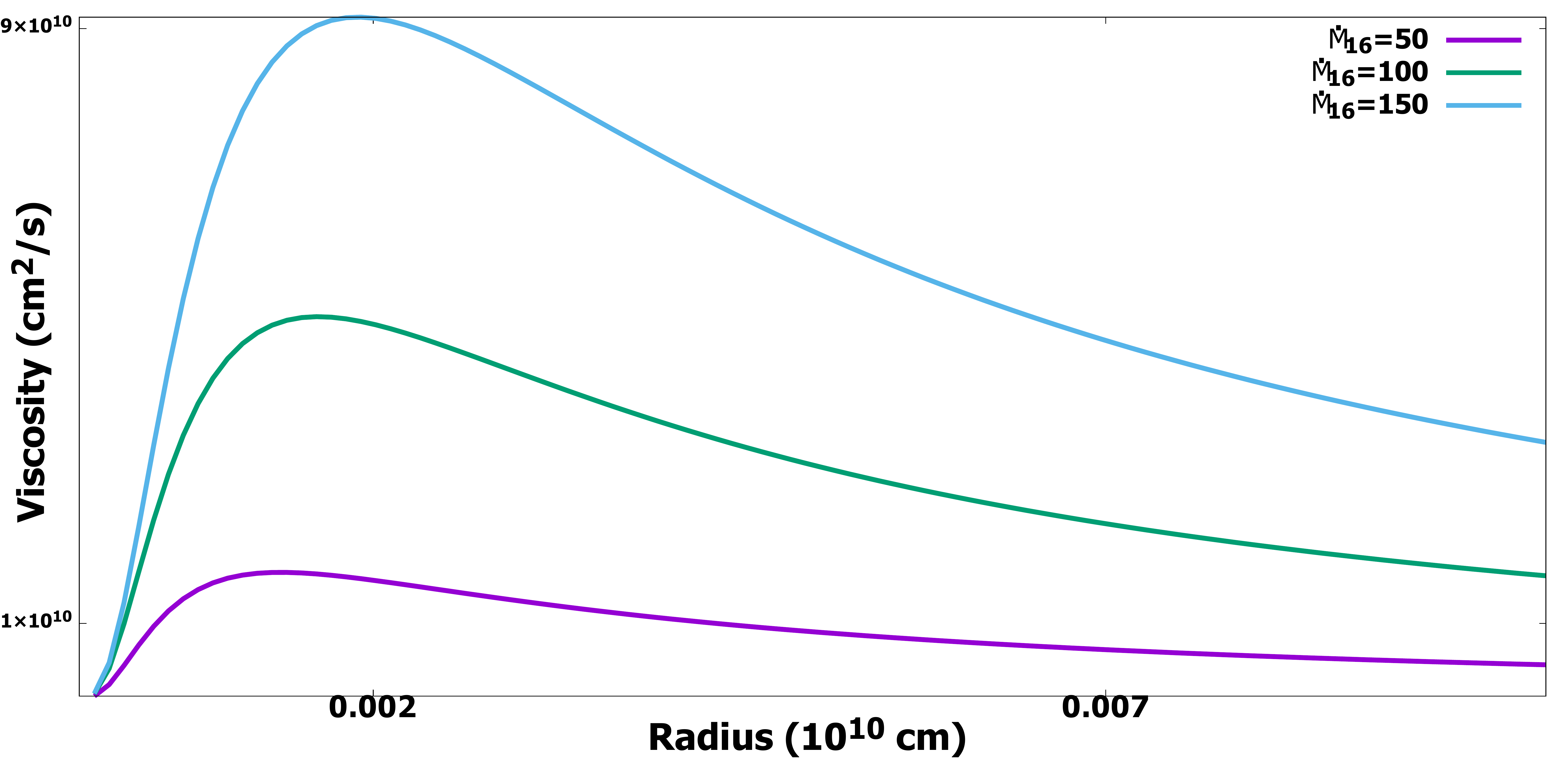}}
\vspace*{8pt}
\caption{Viscosity ($\nu$) vs radius for varying mass accretion rates ($\dot{M}_{16}$) for the cases $\dot{M}_{16} = 50 , 100$ and $150$.\protect\label{f7}}
\end{figure}

\begin{figure}[h]
\centerline{\includegraphics[width=3.2in]{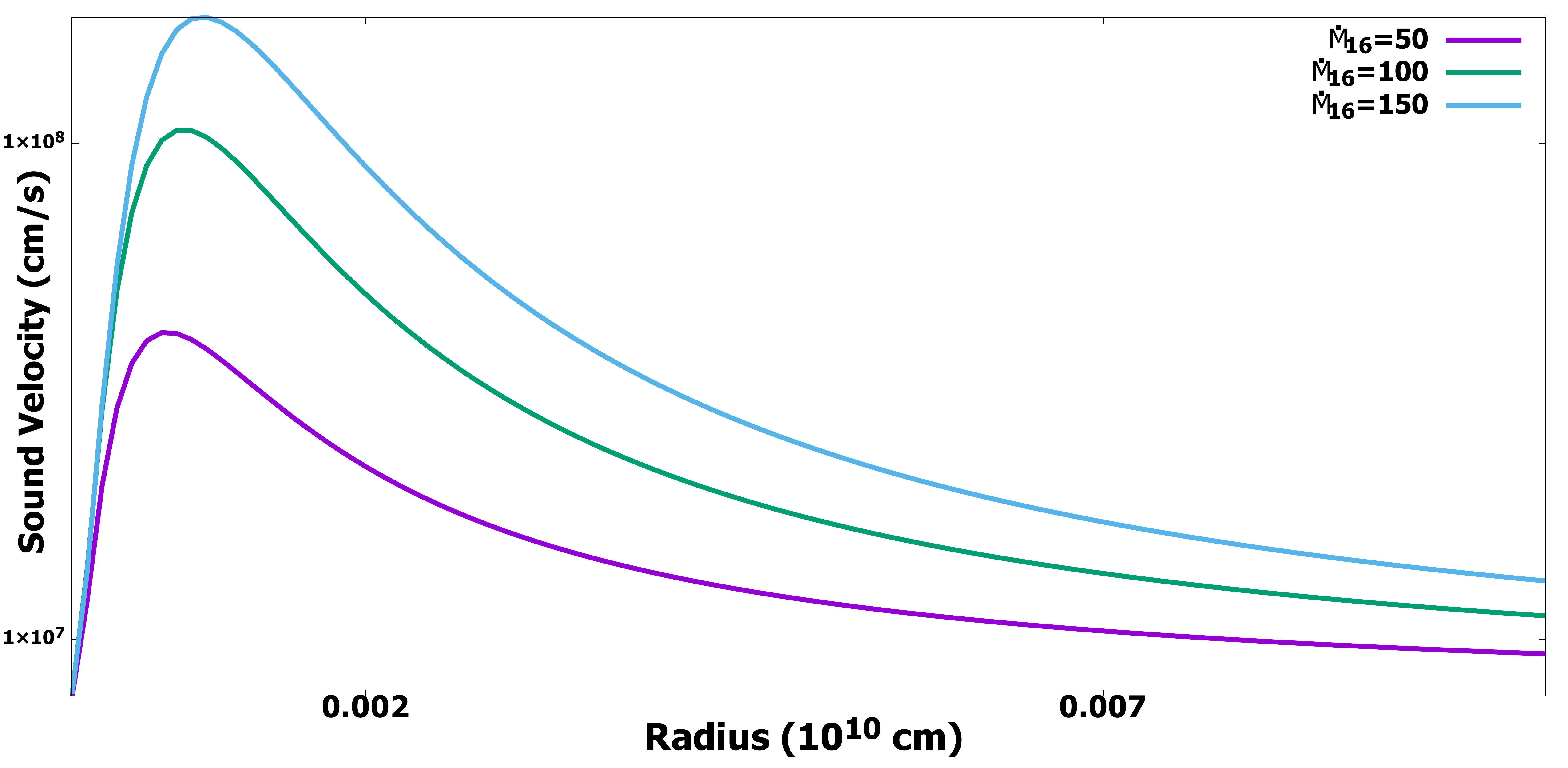}}
\vspace*{8pt}
\caption{Sound speed ($c_{s}$) vs radius for varying mass accretion rates ($\dot{M}_{16}$) for the cases $\dot{M}_{16} = 50 , 100$ and $150$.\protect\label{f8}}
\end{figure}

\begin{figure}[h]
\centerline{\includegraphics[width=3.2in]{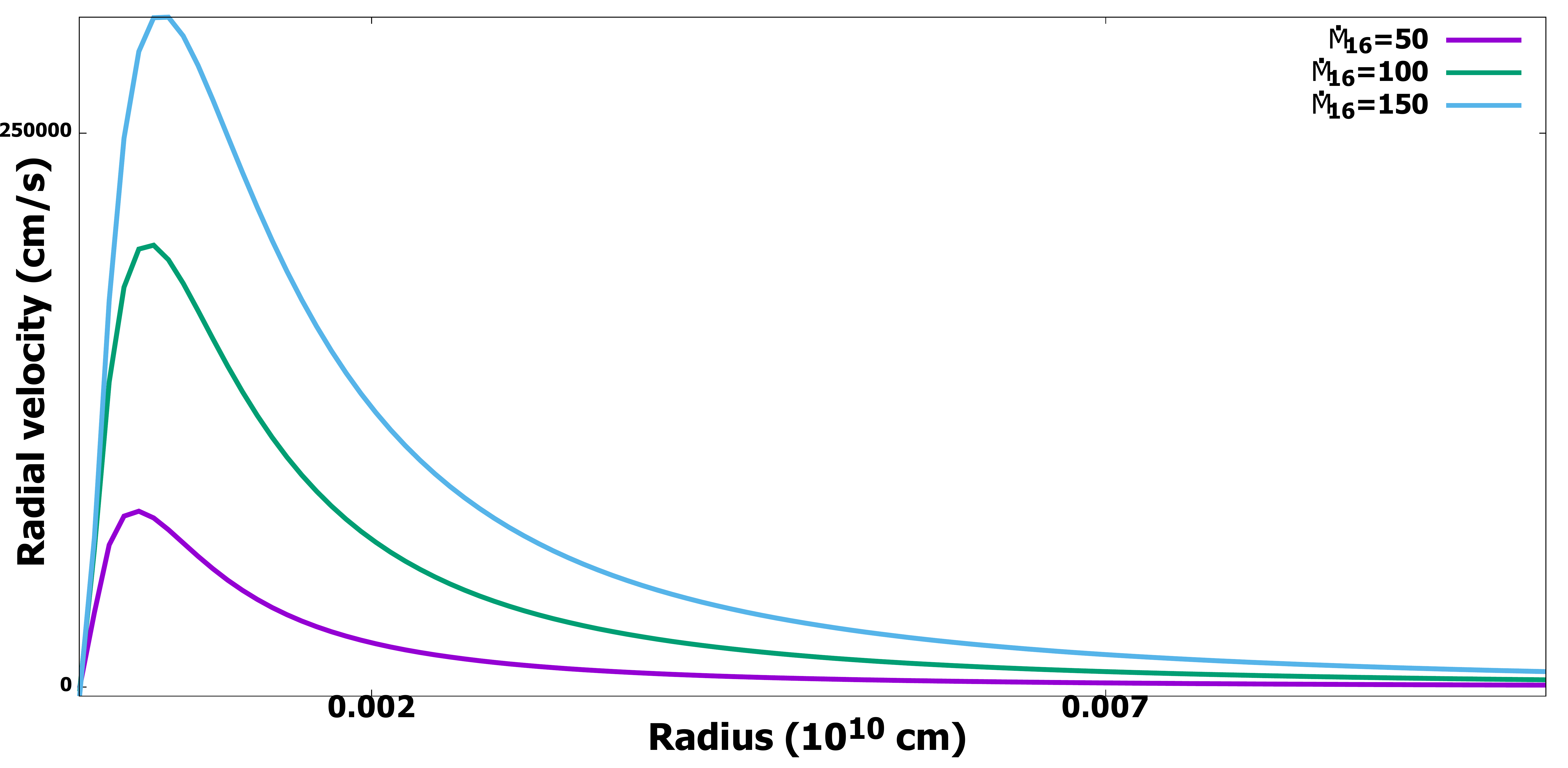}}
\vspace*{8pt}
\caption{Radial velocity ($V_{r}$) vs radius for varying mass accretion rates ($\dot{M}_{16}$) for the cases $\dot{M}_{16} = 50 , 100$ and $150$.\protect\label{f9}}
\end{figure}

Eqs. (\ref{18}) - (\ref{20}) and Eq \ref{22} are now graphically shown in Fig.\eqref{f6} - Fig.\eqref{f9} for varying mass accretion rates ($\dot{M}_{16}$) where we vary $\dot{M}_{16}$ as $\dot{M}_{16} = 50 , 100$ and $150$. Stellar mass is assumed to be $M_{\ast} = 1.4 M_\odot$, radius of the star $R_{\ast} = 10^6 cm$, spin period of the star is set to $0.1$ sec, alpha parameter takes the value $\alpha = 0.01$ and the magnetic field is set to $10^8 G$. The solutions are plotted for $R = 0 - 1000$ km.
Figure \ref{f6} show vertical scale-height as a function of radius and mass accretion rates. From \ref{18}, it is evident that scale-height ($H$) is linearly dependent on $\dot{M}$. $H$ is observed to increase non-linearly for smaller radii. However, further out, $H$ saturates and increase very slowly. \\
Figure \ref{f7} depicts viscosity against radius for varying mass accretion rates where $\nu$ gets amplified rapidly with an increase in $\dot{M}$. This is because viscosity is proportional to the square of the $\dot{M}$ as can be clearly seen from \ref{19}. In the outer gas pressure dominated disc, viscosity is weakly dependent on $\dot{M}$ \cite{matt}). Viscosity is observed to decrease with increasing radii in the inner disc which is in contrast with the profile of viscosity for the outer disc where $\nu$ increases almost linearly with radii \cite{matt}). \\
In Figure \ref{f8}, sound speed $c_{s}$ is shown as a function of radii and mass accretion rates where $c_{s}$ increases linearly with an increase in $\dot{M}$. Profiles of $c_{s}$ are observed to decrease and converge with each other with increasing radii. \\
Figure \ref{f9} shows radial velocity with radius for varying $\dot{M}$. Radial velocity is strongly dependent on $\dot{M}$ evident from \ref{21}. Radial velocity is observed to decrease rapidly with increasing radii.

\section{Boundary Radii}\label{bd}
An accretion disc consists of $3$ distinct regions. These are \cite{ss}\\
A) The inner accretion disc where radiation pressure is dominant $P_{rad}>> p_{gas}$ and opacity is largely due to electron scattering $(\tau_{es})$.\\
B) The intermediate accretion disc where gas pressure is dominant $P_{gas}>>P_{rad}$, electron scattering continues to be the primary source of opacity.\\
C) The outer accretion disc where $P_{gas}>>P_{rad}$ and the opacity is dominated by free free absorption $(\tau_{ff})$.\\
Boundary radius $R_{AB}$ separating region A and B is obtained the equating the radiation and gas pressure in the two regions respectively 
\begin{equation}\label{27}
P_{rad}= 7.6\times10^8 M_{1}^{1/2}R_{10}^{-3/2} \alpha^{-1}
\end{equation}
\begin{equation}\label{28}
P_{gas}= 1.18\times10^5
\alpha^{-9/10}\dot{M_{16}}^{4/5} M_{1}^{7/20}R_{10}^{-51/20}d^{16/5}k^{4/5}  
\end{equation} 
Equating equations ($27$) and ($28$) we get,
\begin{equation}\label{29}
R_{AB}= 23.92 \dot{M_{16}}^{16/21} M_{1}^{-1/7} \alpha^{2/21} d^{64/21} k^{16/21} km  
\end{equation}
From equation (\ref{29}) we note that $R_{AB}$ is weakly dependent of $\alpha$  parameter. Since $k$ is always less than unity for a magnetic star, presence of $k$ shrinks the radiation pressure dominated disc around magnetic stars. For a non-magnetic star $k=1$ and $R_{AB}$ becomes maximum for a given stellar mass ($M_{\ast}$), accretion rate ($\dot{M}$) and spin period ($P_{spin}$). Hence we conclude that for a magnetic star the radiation pressure dominates over gas pressure for a lesser radial extent than in a non-magnetic star with similar stellar parameters ($M_{\ast}$, $\dot{M}$, $P_{spin}$, $\mu$). Thus for very high magnetic fields ($B > 10^{10} G$), the inner region of the disk is cut and the contribution of the magnetic field to the disc becomes limited.   From the equation it is also evident that this region can only exist around accretion discs of black holes and neutron stars and also for high accretion rates $(\dot{M_{16}}>1)$.\\
The radius $R_{BC}$ separating region B and C is obtained by equating the expressions of opacity in the two regions.  
\begin{equation}\label{30}
\kappa_{kramer}=\frac{\tau}{\Sigma}=48.64 \dot{M_{16}}^{-1/2} M_{1}^{1/4} R_{10}^{3/4} d^{-2} k^{-1/2} 
\end{equation} 
\begin{equation}\label{31}
\kappa_{e.s}= \frac{\sigma_{T}}{m_{p}}= 0.4 cm^{2}gm^{-1} 
\end{equation} 
equating equations ($30$) and ($31$) we get,
\begin{equation}\label{32}
R_{BC}= 1.656\times10^2 {\dot{M}_{16}^{2/3}} M_{1}^{-1/3}d^{8/3}k^{2/3}km
\end{equation}
Which is independent of $\alpha$ parameter. From equation (\ref{32}) it is clear that except for very high accretion rates ($\dot{M_{16}}>1$), this region is smaller than the radius of a typical white dwarf. Presence of $k$ further shrinks this region in case of a magnetic star. Hence in accretion discs of white dwarfs and other astrophysical objects (except black holes and neutron stars) Kramers opacity can be valid throughout the entire disc. 
\section{Viscous Timescale}
Since radial structure of a thin disc transmute on viscous timescales we broaden our work by studying the modification of viscous timescale in time-dependent discs around neutron stars.\\
Viscous timescale is defined as \cite{frank}
\begin{equation}\label{33}
t_{vis} \sim\frac{R^{2}}{\nu}
\end{equation}
In radiation pressure dominated discs, the viscous timescale ($t_{vis,rad}$) reads
\begin{equation}\label{34}
t_{vis,rad} \sim 0.34 \times10^{15} \alpha^{-1} \dot{M}_{16}^{-2} M_{1}^{-1/2} R_{10}^{7/2} d^{-8}k^{-2}
\end{equation}
which is inversely proportional to the square of magnetic correction term $k$. Thus in the inner radiation pressure dominated discs, the timescale on which the diffusion of matter take place owing to the effect of the viscous torques increases by many folds as $k<1$ for magnetic stars.\\
In gas pressure dominated discs, viscous timescale ($t_{vis,gas}$) is obtained as 
\begin{equation}\label{35}
t_{vis,gas} \sim 0.34 \times 10^{6} \alpha^{-4/5} \dot{M}_{16} ^{-3/10} M_{1}^{1/4} R_{10}^{11/4}d^{-6/5}k^{-3/10}
\end{equation}
Which is also inversely proportional to $k$, although the dependency is not severe. Thus for gas pressure dominated discs, the magnetic effects does not have significant influence on the viscous or radial drift timescale. \\
Apart from viscous timescale the disc structure get affected on many shorter timescales such as \textit{dynamical timescale} and \textit{thermal timescale}. However we shall not study them here as both of these timescales are independent of magnetic parameter ($k$).

\section{Conclusions}
An analytic solution have been developed following the prescription in \cite{matt} to study the inner accretion disc dominated by radiation pressure, under the influence of a central magnetic field. The solutions are obtained in a similar approach to the non-magnetic Shakura \& Sunyaev \cite{ss} model, and undeniably the profiles tend to the non-magnetic form when magnetic moment $\mu$ of the star tends to zero. The solution have been studied around a typical neutron star and the effects of varying spin periods and mass accretion rate on the disc parameters are analyzed. Our results can be summarized as follows:
\begin{itemize}
\item The vertical scaleheight $H$ is weakly and indirectly dependent on radii and increases slowly with increasing $R$. For the non-magnetic Shakura \& Sunyaev solutions, magnetic parameter $k=1$ and thence $H$ is thickest. The scale-height ($H$) is linearly dependent on $\dot{M}$ and upon varying $\dot{M}$, $H$ increases rapidly for smaller radii but further out, $H$ saturates and increase very slowly.
\item Viscosity is found to decrease substantially with an increase in spin period. Additionally, viscosity for non-magnetic case is several orders of magnitude higher than magnetic cases. Since viscosity is strongly dependent on mass accretion rate, when $\dot{M}$ increases, $\nu$ increases rapidly. It should also be noted that in the outer gas pressure dominated disc, viscosity is weakly dependent on $\dot{M}$.
\item Sound velocity is found to decrease with radii and for an increase in spin period. We also report that $c_{s}$ increases linearly with an increase in $\dot{M}$. 
\item Surface density diminishes with an increase in mass accretion rate and spin period as $ \Sigma \propto \dot{M_{16}}^{-1} k^{-1}$. Hence for two stars with identical stellar parameters, disc becomes thicker for the magnetic star. 
\item Radial velocity decreases with increase in spin period of the star. We also found that for very high accretion rates ($\dot{M}_{16}\geq 1$) and very near to the truncation radius ($R_{t}$), radial velocity for the magnetic cases deviate the most from the non-magnetic Shakura \& Sunyaev profile. Radial velocity for the non-magnetic case is higher than all the magnetic cases.  
 As reported in \cite{matt} the solutions differ because the inner disc gets truncated at a larger distance due to magnetically induced accretion. 
\item Expressions of boundary radii separating distinct disc regions have been obtained in the magnetic case. We report that the correction term $k$ helps to shrink the radiation pressure dominated disc with its effect increasing as magnetic parameter $\beta$ increases. However if mass transfer rate is high it can overpower the magnetic influences of the central object and the disc will revert back to non-magnetic form. The magnetic term $k$ is also involved in reducing the radial extent upto which electron scattering continue to be the dominant source of opacity. Thus for a highly magnetic neutron star (large $\beta$) with low mass accretion rate, Kramers opacity can be valid all throughout the disc as the inner disc is unlikely to be present.
\item We also report necessary modification to viscous timescales ($t_{vis}$) in both inner and outer discs due to the presence of the magnetic correction term. Such modifications increases $t_{vis}$ by many folds in the inner disc while the outer disc regions are not severely affected.
\end{itemize}
As a final note we add that the model presented in this work could be beneficial for studying the inner discs of LXMBs with very high accretion rates coupled with magnetic fields of the order of $B \sim 10^{8} - 10^{10}G$.
\section*{Acknowledgments}
We thank K. Sriram for useful criticism and suggestions. We also thank P.K. Sahoo and Biswajit Pandey for constant support and motivation. Most of the work was done when the author was an undergraduate student at Osmania University. We are very much grateful to the honorable referee for the illuminating suggestions that have significantly improved our work in terms of research quality and presentation.


\end{document}